\documentclass[]{spie}  

\usepackage{amsmath,amsfonts,amssymb}
\usepackage{graphicx}
\usepackage[colorlinks=true, allcolors=blue]{hyperref}

\title{Laboratory characterization 
of Hierarchical Fringe Tracking}

\author[a]{Roxanne Ligi} 
\author[a]{Stéphane Lagarde} 
\author[a]{Romain Petrov} 
\author[a]{Sylvie Robbe-Dubois} 
\author[a]{Julien Dejonghe} 
\author[a,b]{Fatmé Allouche} 
\author[c]{Teng Xu} 
\author[c]{Wei Wei} 
\author[c,d,e]{Massinissa Hadjara} 
\author[f]{Abdelkarim Boskri}

\affil[a]{Université Côte d’Azur, Observatoire de la Côte d’Azur, CNRS, Laboratoire Lagrange, Bd de l'Observatoire, CS 34229, 06304 Nice cedex 4, France}
\affil[b]{European Southern Observatory, Casilla 19001, Santiago 19, Chile
}
\affil[c]{Nanjing Institute of Astronomical Optics $\&$ Technology, National Astronomical Observatories, Chinese Academy of Sciences, Nanjing 210042, China}
\affil[d]{Chinese High Angular Resolution Southern Astronomical Laboratory (CHARSAL), Dept. of Astronomy, Camino El Observatorio $\#$1515, Las Condes, Santiago, Chile. Postal Code: 7591245. Casilla 36-D, Santiago - Chile}
\affil[e]{Astro-Photonics Laboratory/Space and Planetary, Exploration Laboratory (SPEL), Dept. of Electrical Engineering, FCFM, Universidad de Chile, Av. Tupper 2007, Santiago, Chile}
\affil[f]{LPHEA Laboratory, Oukaimeden Observatory, Cadi Ayyad University/FSSM, BP 239, Marrakesh, Morocco}

\authorinfo{Further author information: (Send correspondence to Roxanne Ligi)\\Roxanne Ligi: E-mail: roxanne.ligi@oca.eu}

\begin{document} 
\maketitle

\begin{abstract}
One of the main limitations in long-baseline interferometry lies in its fringe tracking sensitivity. The challenge is therefore to maximize this sensitivity while minimizing the spreading of the signal on the detector. This is the goal at the core of the hierarchical fringe tracking (HFT) concept.
We present the laboratory characterization of the $2^{nd}$ generation HFT chips operating in the near infrared (H band) for up to 4 telescopes with more linear phase and group delay estimators, allowing a strong simplification of the tracking algorithm. 
We show a comparison between theoretical intensity outputs for an optimized phase delay chip, and the laboratory measurements of two chips with different designs. The results do not reach the expectations but get close to them with the 10-outputs chip.
Ultimately, this new chip is intended for implementation on the VLTI, CHARA or on the future Xuyi 100m-baseline Stellar Interferometer using three telescopes.  

\end{abstract}




\section{INTRODUCTION: WHY BUILDING A NEW HFT?}
\label{sec:intro}  

Reaching high magnitude stars has always been a limit in long-baseline interferometry (LBI). The signal is perturbed by the atmospheric piston, which blurs the fringes, leading to short exposures. Hence, less signal reaches the detector, which strongly decreases the number of stars to be observed. 

Fringe trackers are designed to stabilize the piston, hence the fringes. There are several types. The current concept used at VLTI is an ABCD \cite{Lacour2019}\,: it is pairwise chip spreading light between many outputs, e.g., in case of a 4-telescopes fringe tracker, the flux is spread over 24 outputs. However, a minimum of 4 photons per frame per baseline is required to work, which sets the limiting magnitude.     
The hierarchical fringe tracking (HFT) concept is dedicated to overcome this problem. The main principle is to co-phase pairs of beams and combine them hierarchically. The resulting beams are sent to the next stage, where they are again co-phased and combined, and so on. A key advantage of this approach is that the output intensity only depends on the co-phasing quality achieved at the first stage and remains independent of the total number of apertures. Thus, contrary to common ABCD chips, the output flux is packed over fewer pixels, allowing one to observe fainter objects. 


Depending on the designed phase shift, the output flux can be different. The extraction of the photometry is used to calibrate this output, but involves removing a part of the flux. The optimized design is thus a trade off between sensitivity and control. As a matter of fact, previous studies on HFT had already been performed \cite{Allouche2024, Petrov2024} and were consistent with expectations. However, their designs were not optimum.

In this work, we present the laboratory tests of H-band 4T photonic chips that should serve to build the MACADAM project (Petrov et al., SPIE, Paper No. 14148-59). Reaching this goal is crucial to measure the size of Active Galactic Nuclei (AGNs) with sufficient precision that the best constraint of the Hubble constant $H_0$ would be possible through the Broad Line Region (BLR) method\cite{Petrov2022}. 

\section{OPTIMIZATION AND TESTS}


\subsection{Design of the new chip}

Among the different methods to optimize an HFT, one is to optimize the phase delay $\theta$. The previous HFT\cite{Petrov2024} had $\theta =  - \pi /4$ in the first level, which allowed injecting a maximum of flux in the second level, in particular $C^+ = C^-$, so the two outputs could be gathered. However, the visibility-to-pixel matrix (V2PM), which allows to compute the output signals from the input ones and depends on $\theta$, was not reversible. The V2PM matrix is reversible if $\theta \ne - \pi/4$. The phase estimator depended on the input flux ratio, which was not ideal.  
Our now chip has $\theta = 0$, which means that the V2PM is reversible, allowing to find the phase delay, the piston and tracking the fringes.

\begin{figure}
    \centering
    \includegraphics[width=0.52\linewidth]{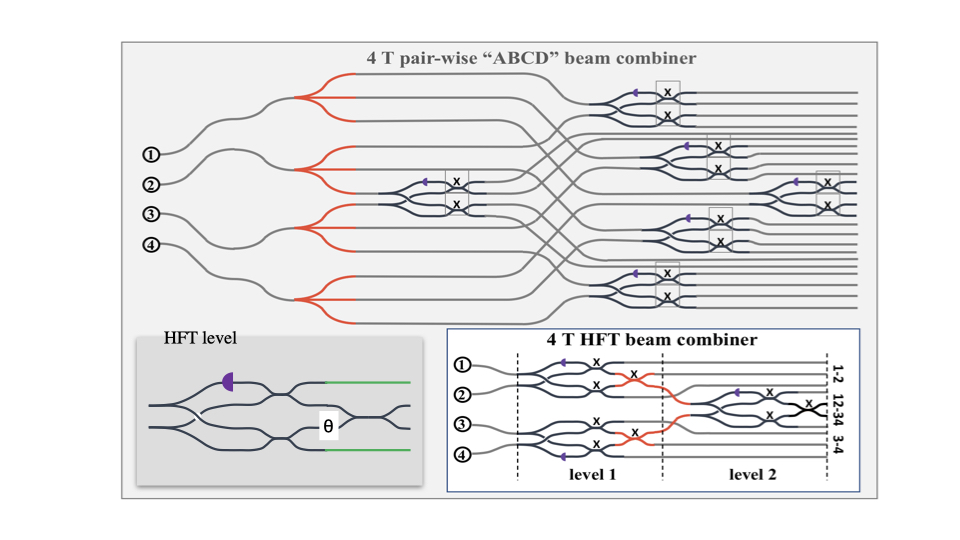}
    \includegraphics[width=0.45\linewidth]{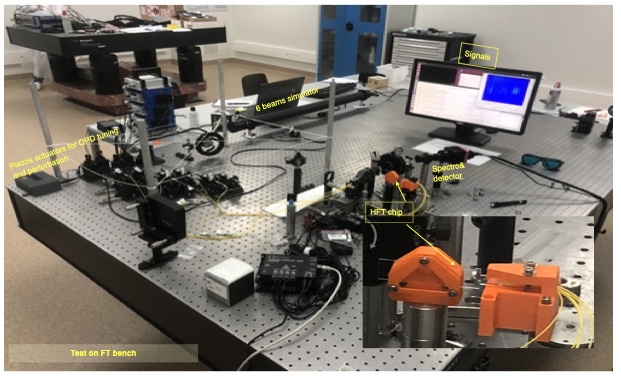}
    \caption{Left\,: comparison between ABCD and HFT chip. Beam splitters are shown in red. Right\,: the test bench. }
    \label{fig:chips}
\end{figure}

\begin{figure}
    \centering
    \includegraphics[width=0.35\linewidth]{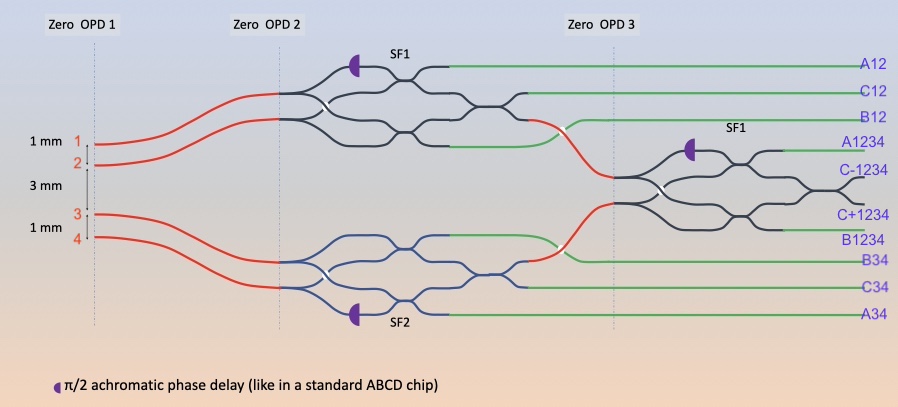}
    \includegraphics[width=0.45\linewidth]{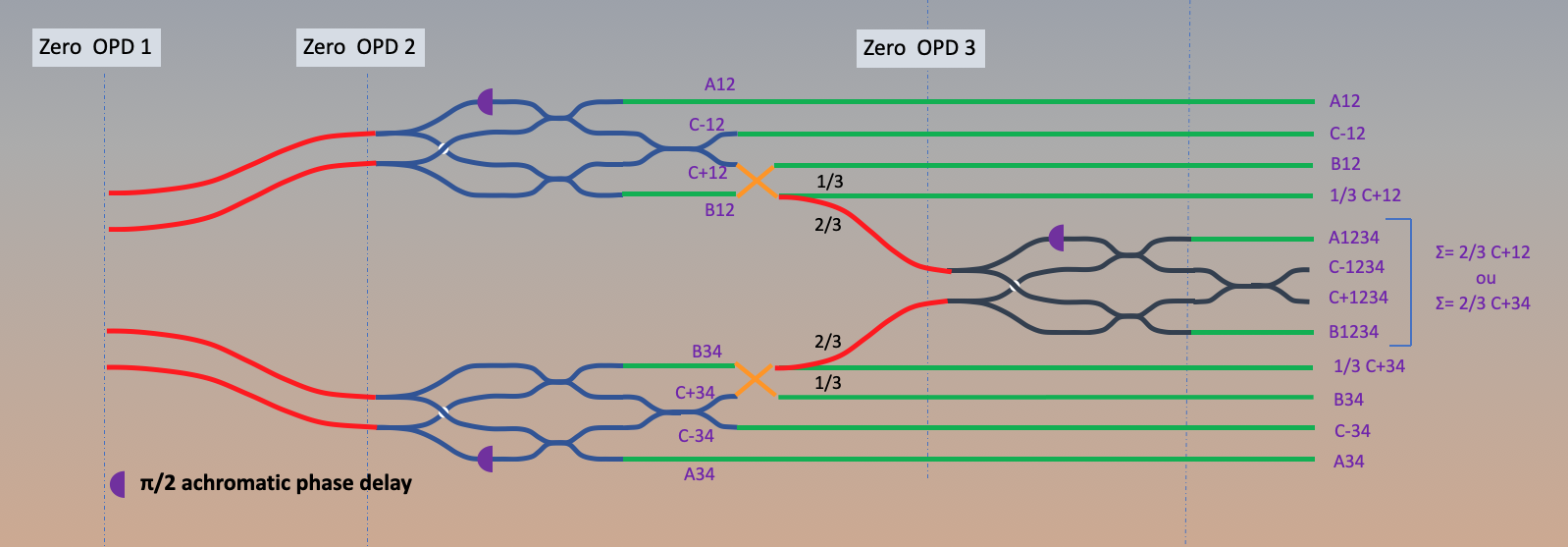}
    \caption{Drawing of the 10-outputs (left) and 12-outputs (right) chip. Crosstalks are drawn in orange. Note that there is no $C^+12$ output in the 10-outputs chip.}
    \label{fig:chipsdrawings}
\end{figure}

In Fig. \ref{fig:chips}, we show a classical ABCD chip compared to a new HFT chip. Each level is composed of one or several cells shown on the left. In this example, there are 10 outputs, but we also tested a 12-output chip (Fig. \ref{fig:chipsdrawings}). However, it is possible to add outputs, for example for photometry. 

\begin{figure}
    \centering
    \includegraphics[width=0.45\linewidth]{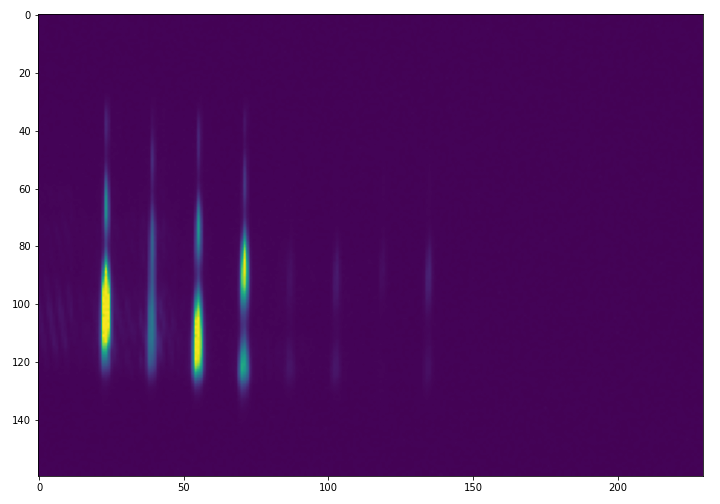}
    \includegraphics[width=0.45\linewidth]{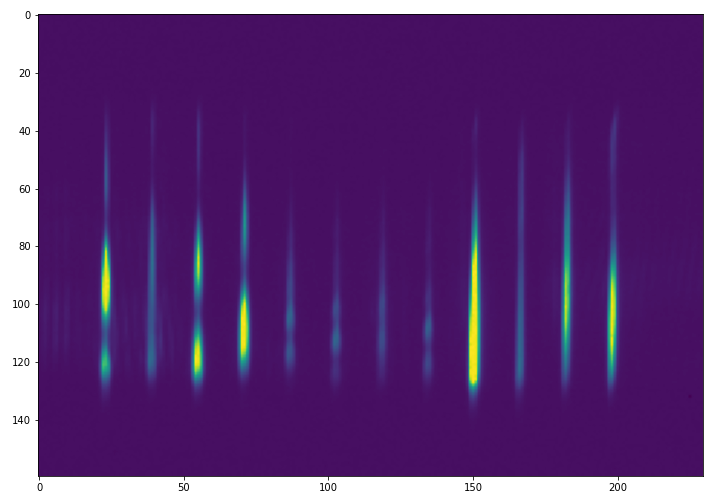}
    \caption{Example of signal on the detector for the 12-outputs chip. Left\,: signal when one pair of entrances are injected. Right\,: signal when all inputs are injected.}
    \label{fig:fluxondetector}
\end{figure}

Our new chips are set up on a test bench in clean rooms at the Lagrange laboratory (Fig. \ref{fig:chips}, right). The flux coming from a collimated infrared source is split into four beams by two sets of mirrors mounted on piezoelectric actuators used either to introduce modulations or to adjust the optical path difference (OPD) between the beams. The light is then injected into optical fibers arranged in a V-groove array, whose outputs are aligned with the entrance of the chip. At the exit of the chip, the light is dispersed by a prism to produce dispersed fringes that are finally recorded by the detector (Fig. \ref{fig:fluxondetector}).

\subsection{Laboratory tests}

The tests were performed over several chips to identify which ones correspond to the expected design and to check if the output intensity is consistent with the theory.

We first need to manually cophase each pair of input by moving the piezoelectric motors. Once it is roughly done, and we found the white fringes, we launch a routine that performs scans by moving the motors, until the fringes are perfectly found. 

We compute, for each peak, the integral of the output light (Fig. \ref{fig:fluxondetector}) after removing the mean of the background noise. The 4 middle outputs intensities can be expressed as:

\[
\begin{pmatrix}
A_m \\
B_m \\
C_m^{+} \\
C_m^{-}
\end{pmatrix}
=
\frac{1}{4}
\begin{pmatrix}
\alpha & \alpha & -2\alpha & 0 \\
\beta & \beta & 0 & 2\beta \\
\gamma_{+}(1+\sin\theta) &
\gamma_{+}(1+\cos\theta) &
\gamma_{+}(1+\cos\theta+\sin\theta) &
-\gamma_{+}(1+\cos\theta+\sin\theta) \\
\gamma_{-}(1-\sin\theta) &
\gamma_{-}(1-\cos\theta) &
\gamma_{-}(1-\cos\theta-\sin\theta) &
-\gamma_{-}(1-\cos\theta-\sin\theta)
\end{pmatrix}
\begin{pmatrix}
I_1 \\
I_2 \\
X \\
Y
\end{pmatrix}.
\]

with $A_m$, $B_m$, $C^+_m$ and $C^-_m$ the measured output intensities, I1, I2, X and Y the input ones, $\theta$ the phase delay and $\alpha, \beta, \gamma_-, \gamma_+$ the transmissions.

The expectation is that the direct output, which do not pass the second stage, should have a lower flux than the central outputs\,; the indirect outputs, bypassing the second stage and following equivalent optical paths, should carry similar flux.

\section{RESULTS}

\begin{figure}
    \centering
    \includegraphics[width=0.6\linewidth]{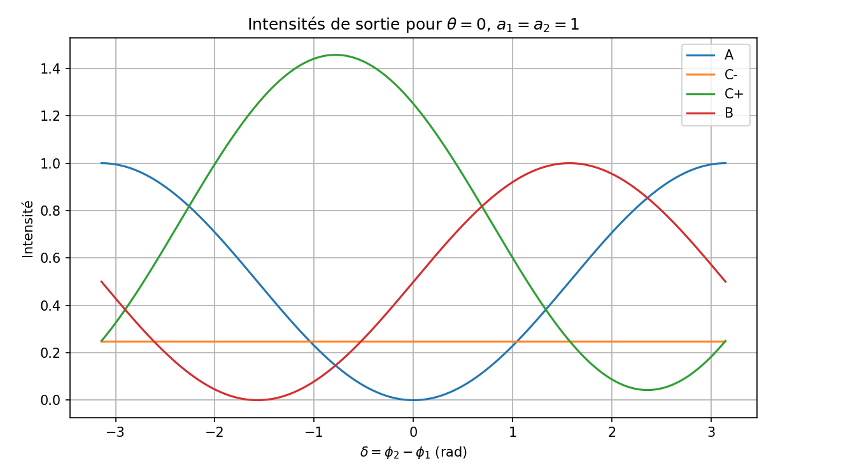}
    \caption{Theoretical intensities for a chip cell with $\theta=0$. This design is optimized to inject a maximum flux in the second level.}
    \label{fig:intensity theory}
\end{figure}

In theory, the output $A12$ should be equal to $B12$ in intensity, $C^-$ should be constant and $C^+$ should have a higher intensity than A and B (Fig. \ref{fig:intensity theory}). This is valid for the case of $\theta=0$, which is more favorable, since it is supposed to inject a maximum flux in the second level. This is the theory considering only one level of the chip. When there is a second level (central outputs), the flux from $C^+$ is supposed to be the sum of the 4 middle outputs.

\begin{figure}
    \centering
    \includegraphics[width=0.45\linewidth]{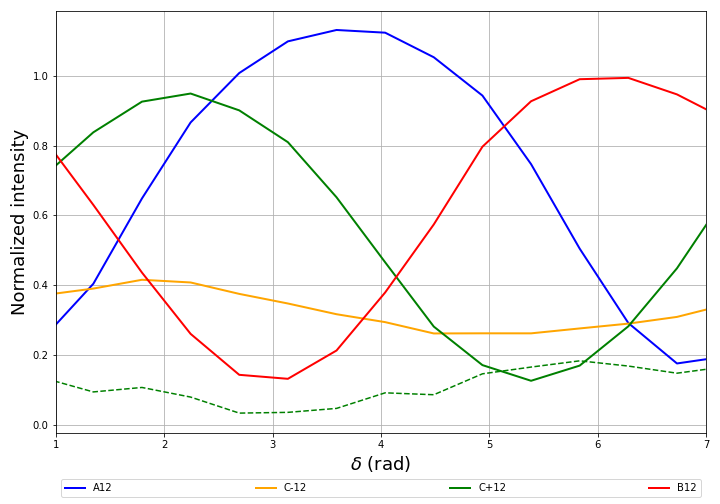}
    \includegraphics[width=0.45\linewidth]{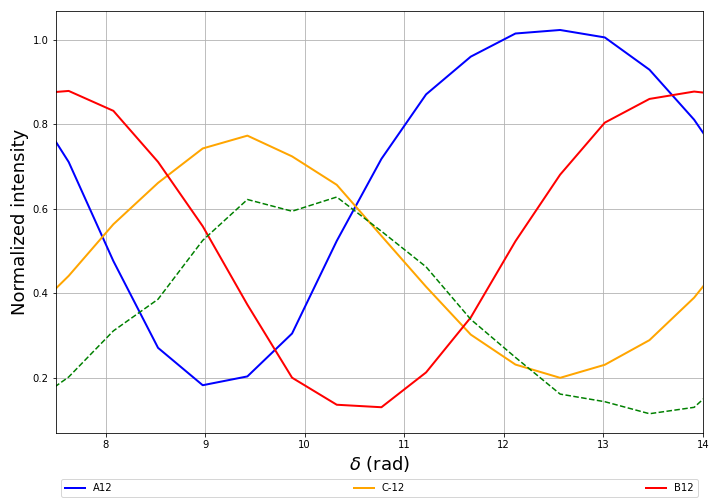}
    \caption{Left\,: normalized intensity for the 12-outputs chip. Only the beams 1 and 2 are injected, so there are four outputs ($A12, B12, C^-12, C^+12$), and the four central outputs. The dashed green line represents the sum of the 4-central outputs, that should be equal to $2C^+12$. Right\,: normalized intensity for the 10-outputs chip. Only beams 1 and 2 are injected. The dashed green line also represents the sum of the 4-central outputs, which is more consistent with expectations. }
    \label{fig:intensity measurements}
\end{figure}

Figure \ref{fig:intensity measurements} shows the output intensity for two different chips, one with 10 outputs and another one with 12 outputs, when light is injected in two inputs only, in order to correspond to the theoretical chip. 

For the 12-outputs chip (Fig. \ref{fig:intensity measurements}, left), we see that fluxes are not correctly phased. This chip includes a photometry output ($C^+12$) dedicated to monitoring the output photometry. The flux going through $C^+$ is supposed to be divided in 1/3 and 2/3 going to the second stage. The dashed green line represents the sum of the four middle outputs. We see that twice this output is not equal to $C^+12$ (plain green line). We thus suppose that the problem comes from the incorrect design of the chip, that does not divide the flux as expected. 

To avoid this problem, we try another chip with 10 outputs, that does not involve such division of the flux. Fig. \ref{fig:intensity measurements} (right) shows that the outputs are more correctly phased. The output intensities between $A12$ and $B12$ are more or less equivalent. The green line represents the sum of the four central outputs, which is however much lower than expected. We also see that the $C^-12$ is not constant as it should, which suggests that too much flux goes in the $C^-12$ path instead of going to the second level. 

In both cases, the output intensity of $A$ and $B$ do not reach $0$ at the lowest. This can be due, at least partially, to a bad removal of the background, or flux loss across the chips. 

\section{CONCLUSION}

We show the results of tests performed on two HFT chips, a 10-outputs one and a 12-outputs one. This latter does not follow the expectations, while the former gets closer to them. Despite that, we know that the previous one with $\theta=-\pi/4$ was working well. We are still investigating the causes of the difference between theory and measurements, and are optimistic that our new HFT chip will soon be optimized, and lead to several phase delays designs ($\theta=\pi/6,$…). We highlight here the difficulties of building a chip that matches the initial designs. 

\acknowledgments 
 
This work was supported by the Action Spécifique Haute Résolution Angulaire (ASHRA) of CNRS/INSU co-funded by CNES.

\bibliography{report} 
\bibliographystyle{spiebib} 

\end{document}